\documentclass[ reprint,superscriptaddress,amsmath,amssymb,aps,pre,showpacs,longbibliography]{revtex4-1}

\usepackage[utf8]{inputenc}
\usepackage{graphicx}
\usepackage{dcolumn}
\usepackage{bm}
\usepackage{xcolor}
\usepackage{tabularx}
\usepackage{amssymb}
\usepackage{pgfplots}
\pgfplotsset{compat=newest}
\usepackage{rotating}
\usepackage{hyperref}
\usepackage{todonotes}
\usepackage{float}

\begin{document}
\title{Light-controlled aggregation and gelation of viologen-based coordination polymers} 

\author {Shagor Chowdhury}
\author {Quentin Reynard-Feytis}
\author {Clément Roizard}
\author {Denis Frath}
\author {Floris Chevallier}
\author {Christophe Bucher}
\email{Corresponding author, christophe.bucher@ens-lyon.fr}
\affiliation{Univ Lyon, Ens de Lyon, CNRS UMR 5182, Laboratoire de Chimie, F69342 Lyon, France}%
\author{Thomas Gibaud}
\email{Corresponding author, thomas.gibaud@ens-lyon.fr}
\affiliation{Univ Lyon, Ens de Lyon, Univ Claude Bernard, CNRS, Laboratoire de Physique, F69342 Lyon, France}%

\date{\today}

\begin{abstract}
Ditopic bis-(triazole/pyridine)viologens are bidentate ligands that self-assemble into coordination polymers. In such photo-responsive materials, light irradiation initiates photo-induced electron transfer to generate $\pi$‐radicals that can self-associate to form $\pi$-dimers. This leads to a cascade of events: processes at the supramolecular scale associated with mechanical and structural transition at the macroscopic scale. By tuning the irradiation power and duration, we evidence the formation of aggregates and gels. Using microscopy, we show that the aggregates are dense polydisperse micron size spindle shaped particles which grow in time. Using microscopy and time resolved micro-rheology, we follow the gelation kinetics which leads to a gel characterized by a correlation length of a few microns and a weak elastic modulus. The analysis of the aggregates and the gel states vouch for an arrested phase separation process.
\end{abstract}

\pacs{xxx}
                             
\maketitle

\section{Introduction}
Stimuli-responsive supramolecular self-assemblies have the potential to organize in various structures depending on the duration and intensity of the stimulus. Those materials may in particular self-assemble into a wide variety of structures ranging from aggregates, capsules, gels or textured thin films~\cite{stuart2010}. The gel state \cite{chivers2019,haring2016,jones2016, panja2021, babu2014, datta2017, dhiman2020, wei2017, xu2018, Zhang2013} is an interesting class of soft materials with broad potential applications ranging from structure directing agents for the synthesis of nanoporous materials, surface patterning, ink, 3D printing to pollutant removal and biomedical applications \cite{okesola2016,kan2017,rodell2015,dong2015,sangeetha2005}.

Supramolecular gels are typically formed in solution from the self-assembly of tailor-made low molecular weight building blocks yielding extended networks stabilized through non-covalent and fully reversible bonds providing the system with elastic properties at rest.\cite{chivers2019,haring2016,jones2016, panja2021, babu2014}. In other words, a sol-gel transition translates molecular information into macroscopic properties. Gelation pathways are not trivial in supramolecular systems since they generally involve a large number of interactions evolving over time~\cite{lan2015}. Moreover, examples where irradiation leads to a strengthening of rheological properties are scarce since photons absorption often results in the collapse of the gel~\cite{draper2016_ChemMater}. In that context, reaching a deep understanding of photo-induced gelation of supramolecular assemblies is crucial for the design of stimuli-responsive soft materials.


Here, we take advantage of the photo-sensitivity of viologen-based supramolecular polymers~\cite{datta2017, dhiman2020, wei2017, xu2018, Zhang2013, li2017, wang2020, correia2019, chowdhury2019, kahlfuss2018} to direct the self-assembly toward aggregates or gels. Forming gels from viologen derivatives is all the more important that such materials are of high interest for various fields such as electrochromism, molecular machines, memory devices, gas storage and separation or energy storage~\cite{ding2019,striepe2017}. In terms of application, reactive gels have already proved useful to the development of electronic such as memristor~\cite{zhang2017}. Moreover, viologens have also been used  as key redox-responsive components in large-area molecular junctions~\cite{han2020, nguyen2018}. All these exploratory works demonstrate the interest of viologen-based material in electronics and  open enormous prospects for innovations.

We have been interested in recent years in the properties of viologen-based supramolecular architectures and coordination polymers grown by spontaneous assembly of organic ligands with various
metal sources~\cite{kahlfuss2016, kahlfuss2019, kahlfuss2021, kahlfuss2018, abdul2018}. In these studies, the ability of viologen derivatives to form $\pi$-dimers \cite{nishinaga2016} in their reduced state has been exploited to achieve a remote control over the organization of molecules within supramolecular assemblies~\cite{kahlfuss2016, kahlfuss2019, kahlfuss2021}. The photo induced reduction of viologen-based acceptors is a well documented phenomenon that has already been used to generate $\pi$-dimers ~\cite{frath2013, yamamoto2018}. We have recently used this approach to achieve a sol/gel transition triggered by light irradiation of the palladium-based coordination polymer~\cite{kahlfuss2018}.

\begin{figure*}
	\centering
    \includegraphics[width=16cm]{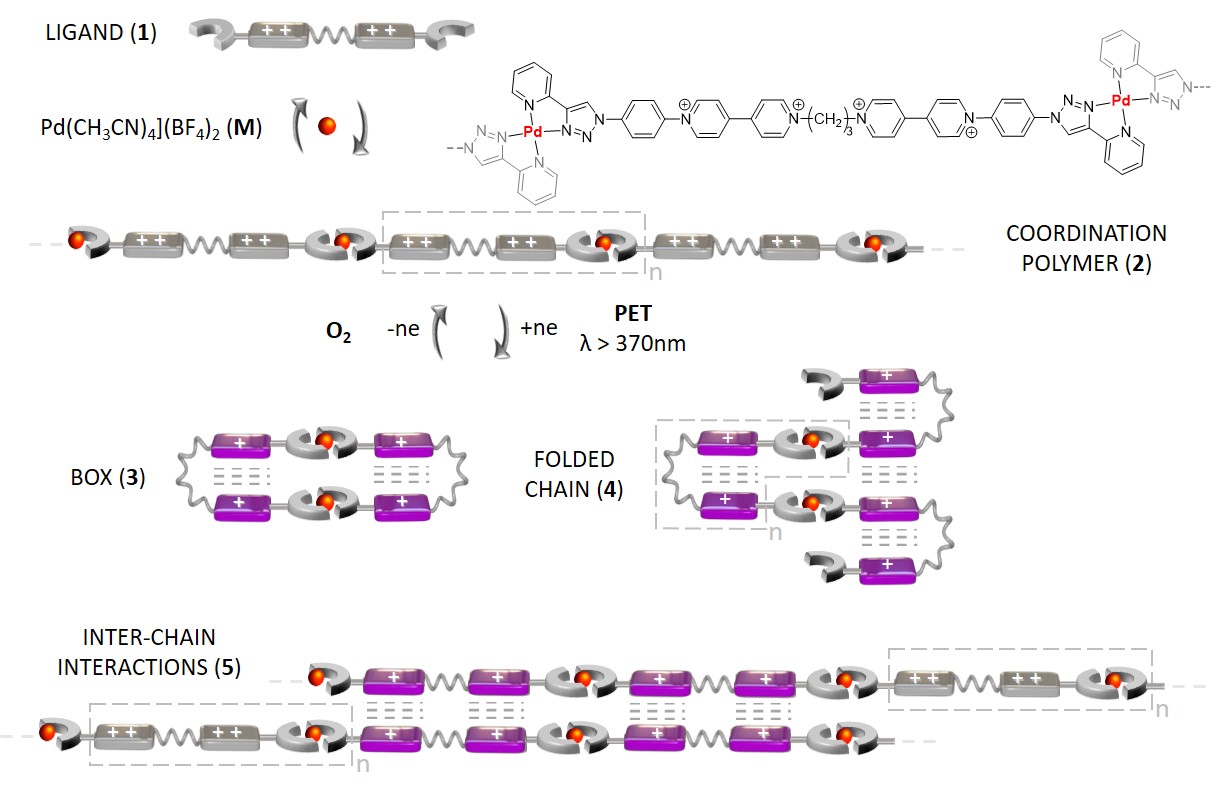}
      \caption{{\bf Schematic representation of the photo-responsive supramolecular coordination polymer used in this study.}
}
    \label{fig:mechanism}
\end{figure*}

As shown in Fig.~\ref{fig:mechanism}, the ditopic ligand \textbf{1} involved in these polymers features two planar nitrogen-containing binding units (triazole-pyridine) introduced on both sides of a viologen-based mechanical hinge that can toggle under stimulation between a folded and a stretched conformation. This ligand was found to self-assemble in the presence of palladium ions to yield linear coordination polymers capable of undergoing large scale reorganizations in solution under light or electrical stimulation~\cite{kahlfuss2018}. The driving force of the folding motion is the $\pi$-dimerisation of the in-situ generated viologen cation radicals. At this stage, the exact structure of the  $\pi$-dimerized structure(s) still remains hypothetical but all available data suggest the formation of either box-shaped discreet complexes (\textbf{3} in Fig.~\ref{fig:mechanism}) or of accordion-shaped folded polymers (\textbf{4} in Fig.~\ref{fig:mechanism}). As mentioned in a recent report, the formation of inter-chain dimers (\textbf{5}  in Fig.~\ref{fig:mechanism}) in the gel phase can also not be ruled out ~\cite{courtois2020}. Following these initial findings, we have carried out detailed microscopy and rheological investigations to provide further insights into the gelation mechanism.


Main goals of the present paper are to map the state diagram of the assemblies obtained after irradiation of the viologen-based coordination polymer and to characterise the kinetics and final properties of the self assembled structures at the microscopic scale. We first describe the preparation of the samples and the characterization techniques that were implemented to perform detailed analyzes on microliters-scale samples under airtight conditions. Second, we present the state diagram of the system depending on the characteristic of the stimulus: a light irradiation of duration $t_i$ and power $P$. As $t_i$ and $P$ increase, we evidence that we can direct the self-assembly toward first aggregates then gels. Third, using microscopy, we show that the aggregates are dense polydisperse micron size spindle shaped particles with an aspect ratio of about two. Fourth, using microscopy and micro-rheology, we show that the gel forms within a few hundreds of seconds. This gel displays a correlation length of about 7 microns and a weak elastic modulus. Those characterisations leads us to formulate a scenario based on arrested phase separation that is compatible with both the aggregates and gel formation.

\section{Materials and methods}

\textbf {Sample preparation.} All samples were prepared under an argon atmosphere (glove box) from compounds obtained according to procedures reported earlier \cite{kahlfuss2018}. In a typical experiment a fresh 100 $\mu$L stock solution was prepared upon mixing equimolar  amounts of \textbf{1} and [Pd(CH\textsubscript{3}CN)\textsubscript{4}](BF\textsubscript{4})\textsubscript{2} in DMF (20 mM) in the presence of triethanolamine (10 mM). The as-prepared solution was introduced into a dismantled quartz absorption cuvette featuring a 0.1 mm layer thickness. The sample was then taken out of the glovebox and submitted to light irradiation during $t_i$ at a power $P$ using a led lamp (X-Cite 120Led Boost from Excelitas) in the reflection mode through the 10x objective of the microscope (Nikon Plan fluor, N.A.=0.3). 

\textbf {Microscopy experiments.} Microscopy experiments were carried out on a Nikon Ti-eclipse inverted light microscope in bright field. We used a 10x and a 40x objective to image the sample. After photo activation, images of the sample were recorded with a camera (ORCA-Flash4 Digital CMOS camera from Hamamatsu) during a time $t$ at 20 Hz. The total recording time was chosen long enough so that the image display stationary statistical properties, $\sim$ 2000~s. The images $I(r)$ are 1024x1024 pixels. The image intensity was encoded in a 16 bits/pixel grey scale. $r$ is the coordinate of a pixel in the image. With the objectives 10x and 40x, one pixel represents respectively 0.644 and 0.161 $\mu$m. 

\textbf {Micro-structure analysis.} Depending on the sample, the structure was analysed either with a shape recognition or a Fourier transform algorithm. For aggregated samples, the molecules form microscopic spindle shape particles. This shape was quantified using an ellipsoid recognition algorithm which gives the diameter of the particle along $r_{\parallel}$ and perpendicular $r_{\perp}$ to its long axis~\cite{modlinska2015}. As set up, this methods is sensitive to structure down to 1.8~$\mu$m. For the gel samples, the molecules form a uniform and isotropic network with a network characteristic length of a few microns. This characteristic length was extracted from the radial average of the Fourier transform ($FFT$) of the image at a time $t$: $I(q,t)=|FFT[I(r,t)-\langle I(r,t) \rangle]| ^2$. The wave vector amplitude $q$ is the variable in Fourier space associated to $r$ in real space. $\langle I(r,t) \rangle$ is the spatial average intensity of the image.    

\textbf {Time resolved micro-rheology.} Micro-rheology measurements involved acquiring movies under the microscope of the sample seeded with colloidal tracers. We used 5 $\mu$m diameter polystyrene colloidal particles (Fluka) as tracers. The amount of colloids was adjusted to reach a diluted colloid regime -- the distance between colloids is at least 5 colloid diameters -- to avoid colloid-colloid interactions. Micro-rheology consists in taking advantage of the dynamics of the embedded colloids. The colloids act as local probes whose dynamics, activated by the thermal motion, depend on the mechanical properties of the sample: the loss modulus $G''$, which accounts for the viscous dissipation of the sample and the elastic modulus $G'$, which accounts for the elastic properties of the sample. To retrieve the viscoelastic properties $G'$ and $G''$ of the sample, micro-rheology experiments requires image analysis and calculations. In brief, image analysis consists in tracking the trajectories $r(t)$ of each embedded colloids in the sample as a function of the time $t$ \cite{maurer2014, parthasarathy2012}. We obtain about 50 trajectories per movies. The tracking algorithm precision on $r(t)$ is about 1/10$^{th}$ of a pixel: $\simeq 16$~nm, with the 40x objective. From those trajectories, the mean square displacement of the colloids $\langle \Delta r^2 \rangle$ is calculated as a function of the lag time $\Delta t$.
Micro-rheology principles \cite{mason1997, gao2007} then rely on computing the complex viscoelastic modulus $G$ or the creep compliance $J$. $G$ is obtained from the unilateral Laplace transform $\mathcal{L}[.]$ of $\langle \Delta r^2 \rangle$ using the generalized Stokes-Einstein equation: $G(s)=\frac{k_BT}{i s a \mathcal{L}[\langle \Delta r^2 \rangle]}$. $i$ is the imaginary number such that $i^2=-1$, $s$ is the Laplace frequency, $k_B$ is Boltzmann's constant, $T=20^{\circ}$C is the temperature of the sample (in Kelvin for the formula) and $a$ is the radius of the colloids. This equation is based on a generalized Langevin equation which describes the sphere's motion in the continuum (neglecting the sphere's inertia), and it is consistent with energy equipartition and the fluctuation dissipation theorem. $G=G'+iG''$ is the shear modulus. Its real part yields the elastic modulus $G'(f)$ and its imaginary part gives the loss modulus $G''(f)$ where $f$ is the frequency in Hz, with $s=2\pi f$. The calculation method is based on \cite{mason1997, gao2007}. An alternative consists in working directly on the mean square displacement which is proportional to the creep compliance $J(t)=k_BT/[\pi a \langle \Delta r^2 \rangle]$~\cite{gittes1997,palmer1998}. $J(t)$ relates the material strain deformation $\gamma(t)$ to the rate of applied stress $\dot{\sigma}(t)$ via $\gamma(t)=\int_{0}^{t}J(t-t')\dot{\sigma}(t)dt'$~\cite{ferry1980}. The direct proportionality of the mean square displacement and creep compliance avoids artifacts from converting the data to the complex viscoelastic modulus $G$~\cite{palmer1998}. Those artefacts can be seen in Fig.~\ref{fig:microrheo}c where $G'$ may drop drastically at the edges of frequency domain. 

As our samples evolve in time, we decomposed the movies in a succession of sequences of duration 50~s. 50~s therefore represents the time during which a micro-rheology experiment is performed and the kinetics resolution. Doing so, we assumed that the sample properties are stationary or evolve very little during this 50~s time scale. 

\section{Results}

\subsection{Directing the microscopic state of self-assemblies with light irradiation}

\begin{figure}
	\centering
    \includegraphics[width=8cm]{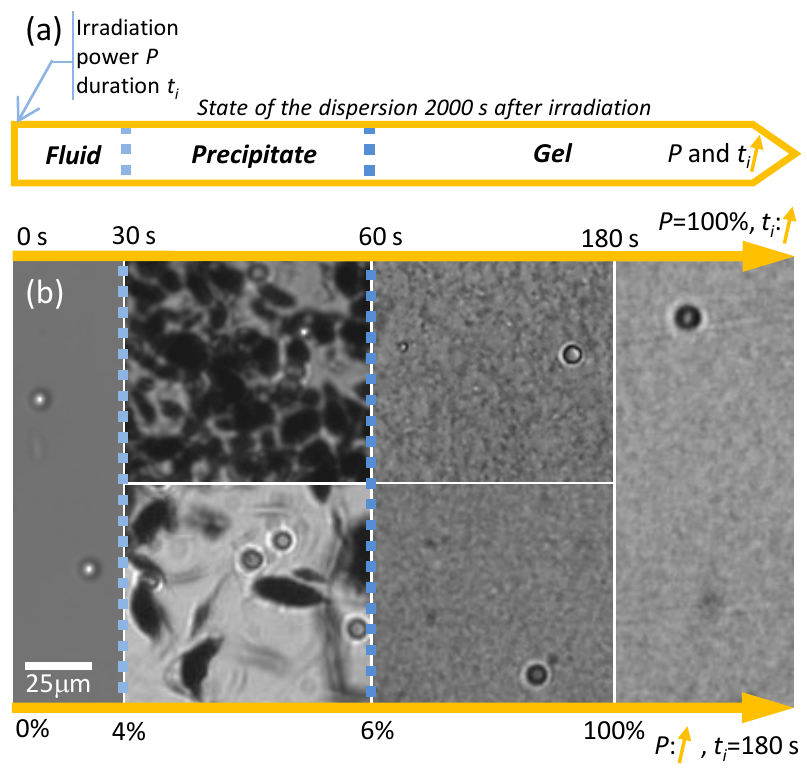}
      \caption{{\bf State diagram as a function of the initial photo-activation time $t_i$ and power $P$.} (a) State diagram of the dispersion inferred from microscopy images and micro-rheology as $P$ and $t_i$ increase. (b) Microscopy images of the sample after reaching its final state at $t=2000$~s after irradiation as function of the photo activation time $t_i$ and the power $P$. Upper arrow: $P=100$\% while $t_i$ varies from 0 to 180 s. Lower arrow: $t_i=180$ s while $P$ varies from 0 to 100\%. Circular spot on the images are the 5 $\mu$m diameter tracer colloids used for micro-rheology.
}
    \label{fig:statediag}
\end{figure}

\begin{figure}
	\centering
    \includegraphics[width=8cm]{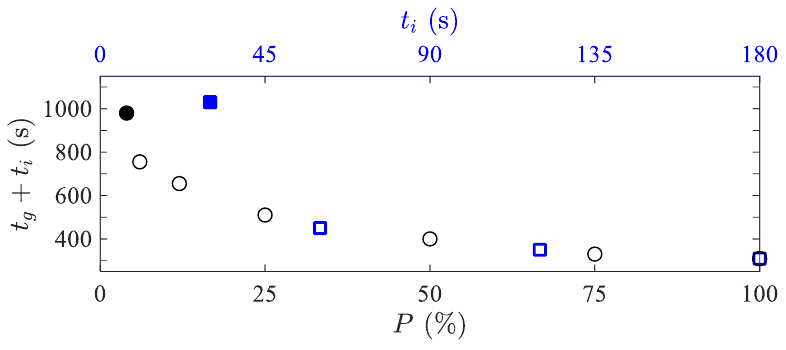}
      \caption{{\bf Aggregation and gelation time $t^*$ as a function of the initial photo-activation time $t_i$ and power $P$.} $t^*$ is define as the first time structures appear on the microscopy images after the irradiation in turn off. Evolution of $t^*+t_i$ as function of $P$ while $t_i$ is kept constant and equal to 180~s (black). Evolution of $t^*+t_i$ as function of $t_i$ while $P$ is kept constant and equal to 100\% (blue). Full symbols correspond to the aggregate state while open symbols correspond to the gel state. 
}
    \label{fig:tg}
\end{figure}

We used microscopy to image the sample and map the state diagram of the dispersion under different irradiation conditions. We varied the irradiation time $t_i$ from 0 to 180~s and the lamp power $P$ from 0 to 100\%. Fig.~\ref{fig:statediag} shows bright field microscopy images of the same sample under different irradiation conditions. Images were taken when the structure of the sample no longer varied, $t = 2000$~s after switching off the lamp. At low $t_i$ and $P$, the sample was found to form aggregates. The aggregated sample is spatially heterogeneous and composed of individual elongated aggregates randomly oriented of size a few tens of microns. At higher $t_i$ and $P$, the sample forms a gel. All the gels display the same structural and mechanical properties: a spatially homogeneous and weakly contrasted network with a correlation length $\xi \simeq 7$~$\mu$m and a weak elastic modulus $G' \simeq 0.3$~Pa. The measurements of those properties are presented later in the article. We define $t^*$ the time when the first structures appear under the microscope, Fig.~\ref{fig:tg}. For aggregates, $t^*$ is of about 1000~s. For gels $t^*$ is lower and decreases with $t_i$ and $P$.

\subsection{Micro-structure characterization of the aggregation and the gelation}

\begin{figure}
	\centering
    \includegraphics[width=8cm]{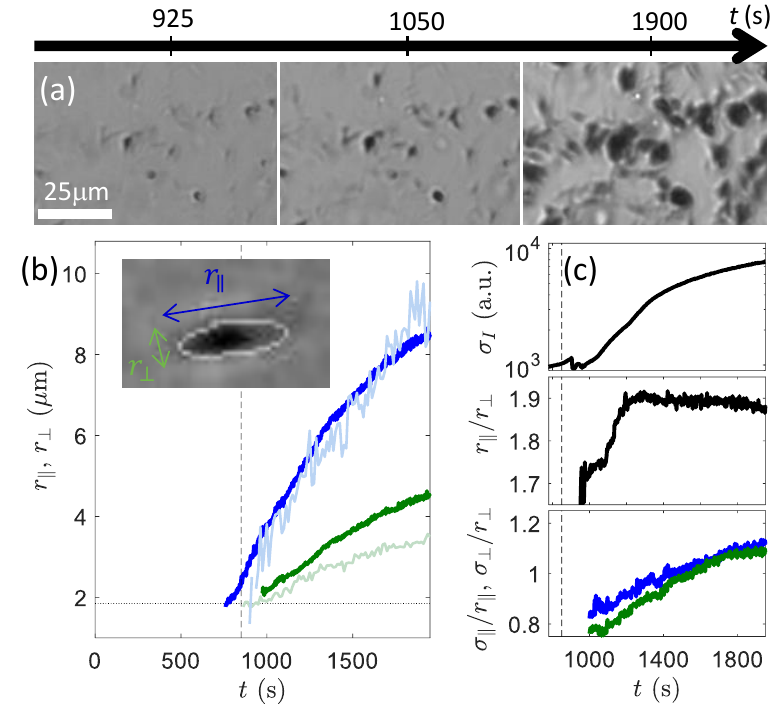}
      \caption{{\bf Aggregation kinetics using time resolved microscopy for an irradiation of $P=100$~\% during $t_i=30$~s} (a) Bright field images of the aggregate $I(r,t)$ at $t=370$, 420 and 700~s. (b) Evolution of the average value of $r_{\parallel}$ (blue) and $r_{\perp}$ (green) as a function of $t$. Light blue and green curves shown the evolution of a typical single particle. Inset: superposition of the image of a particle and its contour. (c) From top to bottom, time evolution of $\sigma_I$ the standard deviation of $I(r,t)$, the particle aspect ratio $r_{\parallel}$/$r_{\perp}$, and the polydispersity. The vertical dash line indicate $t^*=850$~s.
}
    \label{fig:sprecip}
\end{figure}

It is remarkable that a simple change in the time and power of irradiation lead to such different states: aggregate or gel. Irradiation conditions to form aggregates were best achieved using a combination of low power or short irradiation time as indicated in Fig.~\ref{fig:statediag}. As shown in Fig.~\ref{fig:sprecip}a, molecular dispersion irradiated with $P=100$~\% for $t_i=30$~s,gave isolated elongated micro-particles. Although those micro-particles coalesce, within our experimental time window, we never observed the formation of homogeneous phases. The first particle observed under the microscope appears around $t\simeq 850$~s after irradiation. Such particles display two important characteristics. ($i$) A spindle like morphology with a long axis of length $r_{\parallel}$ and a short axis of length $r_{\perp}$. As shown in Fig.~\ref{fig:sprecip}b, on average, $\langle r_{\parallel} \rangle$ and $\langle r_{\parallel} \rangle$ grow with time up to roughly 10 $\mu$m. The average aggregated particles aspect ratio, defined by  $r_{\parallel}/r_{\perp}$, is about 1.8, Fig.~\ref{fig:sprecip}c. The particle polydispersity, defined by the ratio between the standard deviation of the particle size divided by the particle size, is always large and increases from $\simeq 0.8$ to 1.1. ($ii$) High molecular densities, as revealed by the dark coloration of the microscopy images, Fig.~\ref{fig:sprecip}a. The density difference between the particle and the background can be estimated through $\sigma_I$ the standard deviation of the image $I(r,t)$, Fig.~\ref{fig:sprecip}b. Particles are much denser that the gel backbone as shown in Fig.~\ref{fig:statediag}.

\begin{figure}
	\centering
    \includegraphics[width=8cm]{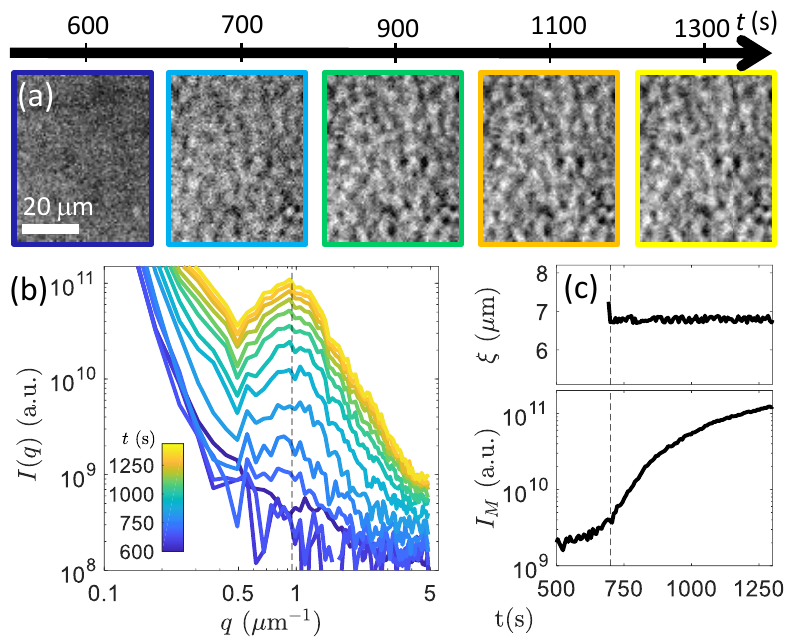}
      \caption{{\bf Gelation kinetics using time resolved microscopy for an irradiation of $P=100$~\% during $t_i=120$~s}. (a) Bright field images $I(r,t)$ of the gelation process at different time $t$. (b) Evolution the Fourier spectrum $I(q,t)$ of the microscope images $I(r,t)$ as function of the wave number $q$ for different time $t$. Colors code for the time $t$ after irradiation. (c) From top to bottom, time evolution of the correlation length $\xi$ and peak intensity $I_M$. The vertical dash line indicate $t^*=690$~s.
      }
    \label{fig:sgel}
\end{figure}

The best gelation conditions were obtained using high power and long duration irradiations (Fig.~\ref{fig:statediag}). As shown in Fig.~\ref{fig:sgel}, with pictures recorded for a molecule
dispersion subjected to irradiation of $P=100$~\% for $t_i=120$~s, the first structures appear under the microscope at about $t=690$~s. The gel kinetics and micro-structure are very different from the aggregation state. First, the gel structure does not shown any evidence of local anisotropy. The images display a well defined length scale. This is evidenced in Fig.~\ref{fig:sgel}b through the Fourier spectrum $I(q,t)$ of the gelling molecular dispersion. $I(q,t)$ shows in particular a peak at $q_p$ of intensity $I_p$. $q_p$ is related to the correlation the length of the gel $\xi$, i.e., the characteristic length of the gel network defined by $\xi=2\pi/q_p$. $I_p$ is an indicator of the structure contrast and therefore the molecular density of the network structure with respect to the background. Above $t=690$~s, $\xi$ is constant, $\xi = 6.9\pm 0.2$~$\mu$m, and $I_p$ increases by almost two orders of magnitude. Contrary to the aggregate state, the gel length scale is set a the early beginning of the kinetics and remains afterward unchanged. However, the contrast and therefore the gel backbone becomes denser during the kinetics. To sum up, the gel forms an homogeneous porous network structure of typical size $\xi \simeq 7$~$\mu$m.

\subsection{Viscoelastic characterization of the sol-gel transition}

\begin{figure*}
	\centering
    \includegraphics[width=18cm]{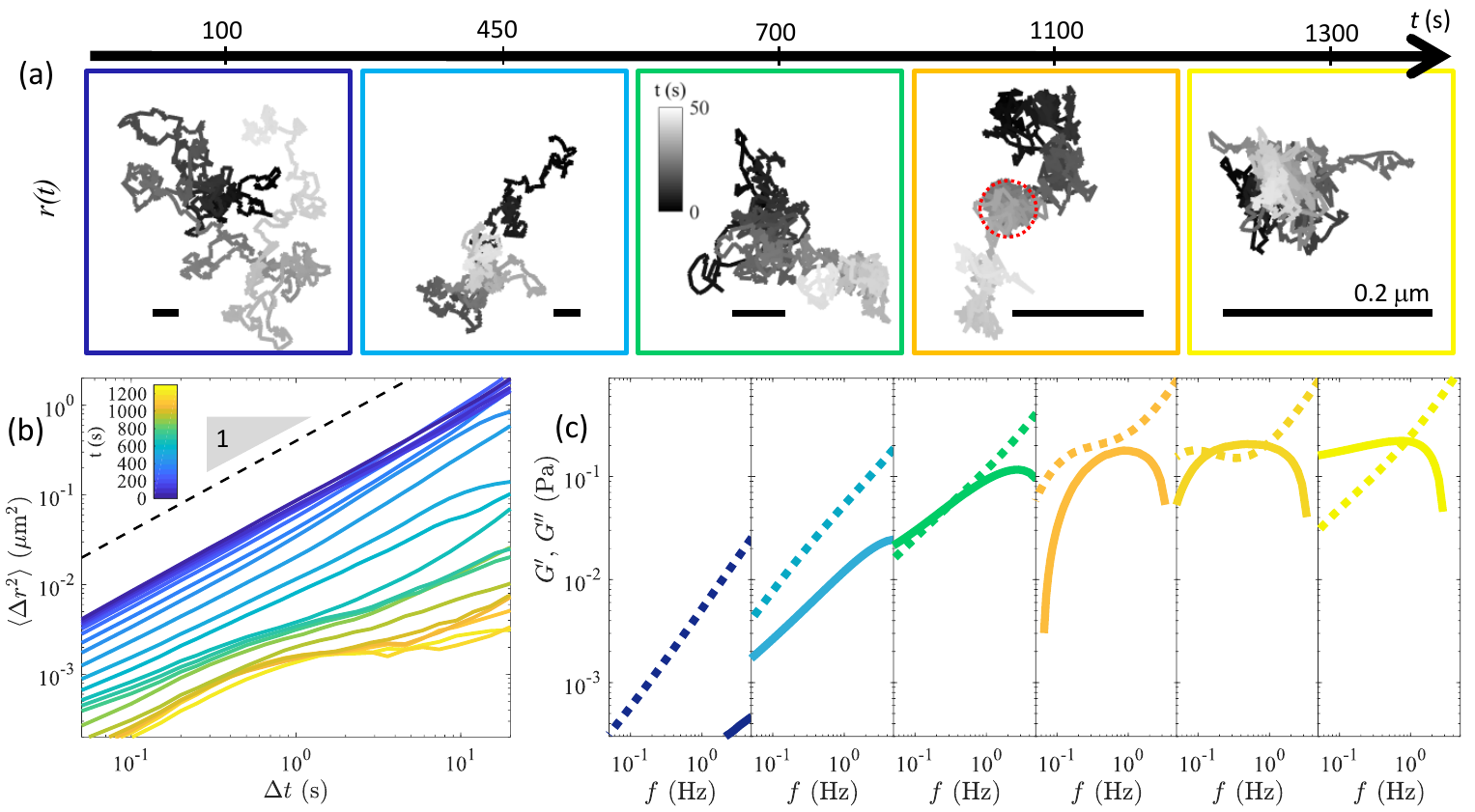}
      \caption{{\bf Gelation -- from the embedded colloids dynamics to time resolved micro-rheology.} This sample is photo-activated during $t_i=120$~s at $P=100\%$. (a) Representative trajectories $r(t)$ of a colloid embedded in the sample at different time $t$ after photo activation. The color from blue to yellow codes for the time $t$ after photo activation. The grey scale codes for the time from 0 s (dark grey) to 50 s (light grey). The scale bar always represent 0.2$\mu$m. Note that the scale bar increases size as $t$ increases, meaning the trajectory is zoom in as $t$ increases. The doted red circle in the trajectory at $t=1100$~s indicate the localization of the colloid before diffusing to other location. (b) Mean square displacement of the colloids $\langle \Delta r^2 \rangle$ as a function of the lag time $\Delta t$. Each $\langle \Delta r^2 \rangle$ is extracted from movies sequences of duration 50~s. (c) Viscoelastic proprieties of the sample deduced from $\langle \Delta r^2 \rangle$: loss $G'$ (line) and elastic modulus $G''$ (dot line) as a function of the frequency $f$.
    }
    \label{fig:microrheo}
\end{figure*}

We have next  focused on the mechanical properties of the molecular dispersions leading to the gel state. We found that irradiation, in the limit where $P$ or $t_i$ is high, induces full gelation of the sample on times scales of the order of $t^*$. In Fig.~\ref{fig:microrheo}, for the gelation conditions, $P=100$~\% during $t_i=120$~s, conditions identical to the sample whose structure is analysed in Fig.~\ref{fig:sgel}, we extract, thanks to the dynamics of colloids dispersed in the sample, the mechanical properties of the sample after photo-activation using time resolved micro-rheology. This analysis is decomposed in tree main steps: tracking the trajectory of the colloids $r(t)$ (Fig.~\ref{fig:microrheo}a), computing mean square displacement of the colloids $\langle \Delta r^2 \rangle(\Delta t)$ (Fig.~\ref{fig:microrheo}b), and extracting the variation of the loss $G''$ and elastic $G'$ modulus as a function the frequency $f$ (Fig.~\ref{fig:microrheo}c). We choose to focus on four characteristic times of the gelation kinetics to describe its evolution: $t=100$, 700, 1100 and 1300~s.

At $t=100$~s, the sample is translucent and homogeneous (Fig.~\ref{fig:sgel}a). As shown in Fig.~\ref{fig:microrheo}a, the colloid trajectory is open and the colloid diffuses micrometric distances during this acquisition of 50~s. As shown in Fig.~\ref{fig:microrheo}b, this corresponds to a linear variation of the mean square displacement $\langle \Delta r^2 \rangle$ with the lag time $\Delta t$: $\langle \Delta r^2 \rangle=4D \Delta t$. The colloid diffuses with a diffusion constant $D=0.018 \mu$m$^2$/s. At this stage, the sample is a newtonian fluid. In, Fig.~\ref{fig:microrheo}c $G'$ is negligible and $G''$ varies linearly with $f$: $G''=\eta 2\pi f$ with $\eta\simeq5$~mPa.s the viscosity of the sample.

At $t=700$ s, the sample display for the first time a structure at the micro scale: a network of correlation length $\xi\simeq 7$ $\mu$m (Fig.~\ref{fig:sgel}a). The colloid trajectory $r(t)$ is more compact (Fig.~\ref{fig:microrheo}a). The mean square displacement behave as a power law of time $\langle \Delta r^2 \rangle \sim \Delta t^{0.6}$. $G'$ and $G''$ are power law of the frequency, parallel and equal to one another. This behavior is very close to the behavior of a dispersion at the gel point as defined by Chambon and Winter~\cite{chambon1987}. We note that the time at which the dispersion structure at the microscale $t^*$ coincides with the gelation time $t_g$ defined by Chambon and Winter, $t_g=t^*$. Given that $t_g$ is measured with a precision of 50~s and $t^*$ of one second, we selected the structure measurements rather that the microrheology measurement to define $t_g$. 

At $t=$1100~s, the structure is very similar to the one observed at 700~s. However, the tracers dynamics become different. Indeed, the colloid's trajectories $r(t)$ display dynamical heterogeneities: a colloid spends some time at a specific location before jumping to another location. In other words, a colloid is mildly trap by the network and explore a small region for a characteristic time before the sample rearrange and let the colloid escape to another location where it is again trapped, and so on. The associated $\langle \Delta r^2 \rangle$ shows at small $\Delta t$ a diffusive behavior, Fig.~\ref{fig:microrheo}b. At intermediate $\Delta t \simeq 1$~s, $\langle \Delta r^2 \rangle$ reaches a plateau around $2.10^{-3}$~$\mu$m$^2$: the colloid is confined to region of average size of $0.045~\mu$m. At long $\Delta t$, the colloid diffuse again. This scenario is typical of viscoelastic fluids \cite{stokes2008} as corroborated by Fig.~\ref{fig:microrheo}c. At high frequencies (short $\Delta t$), $G''>G'$ and $G''\sim f$ the sample is fluid and the energy is dissipated in the thermal vibration of the sample. At intermediate frequencies (intermediate $\Delta t$) $G'\sim G''$ the sample is mildly elastic. At small frequencies (large $\Delta t$), $G'< G''$, the sample can rearrange itself and is therefore fluid.

At $t=1300$~s, the sample kept its micro structure, with a correlation length $\xi\simeq 7$ $\mu$m but the network is more contrasted, Fig.~\ref{fig:sgel}b. The colloid trajectory is now very compact: the colloid is trapped to a single location by the network, Fig.~\ref{fig:microrheo}a. The associated $\langle \Delta r^2 \rangle$ shows a diffusion process at short time scales and then a plateau at long time scales, Fig.~\ref{fig:microrheo}b. This behavior is typical of viscoelastic solids \cite{stokes2008,Gibaud2020}, Fig.~\ref{fig:microrheo}d. At high frequencies $f$, just like the viscoelastic fluid, $G''>G'$, the sample dissipate energy due to the thermal vibrations of the network against the background solvent, but at intermediate and small frequencies $f$, $G'$ dominates $G''$ and $G'$ plateaus to $G'\simeq 0.3$~Pa: The sample cannot rearrange on long time scales and is as such an elastic solid. $G'\simeq 0.3$~Pa is a very low value of elasticity. For such a value, the gel would probably flow due to its own weight if put in an incline vial. 

\begin{figure}
	\centering
    \includegraphics[width=8cm]{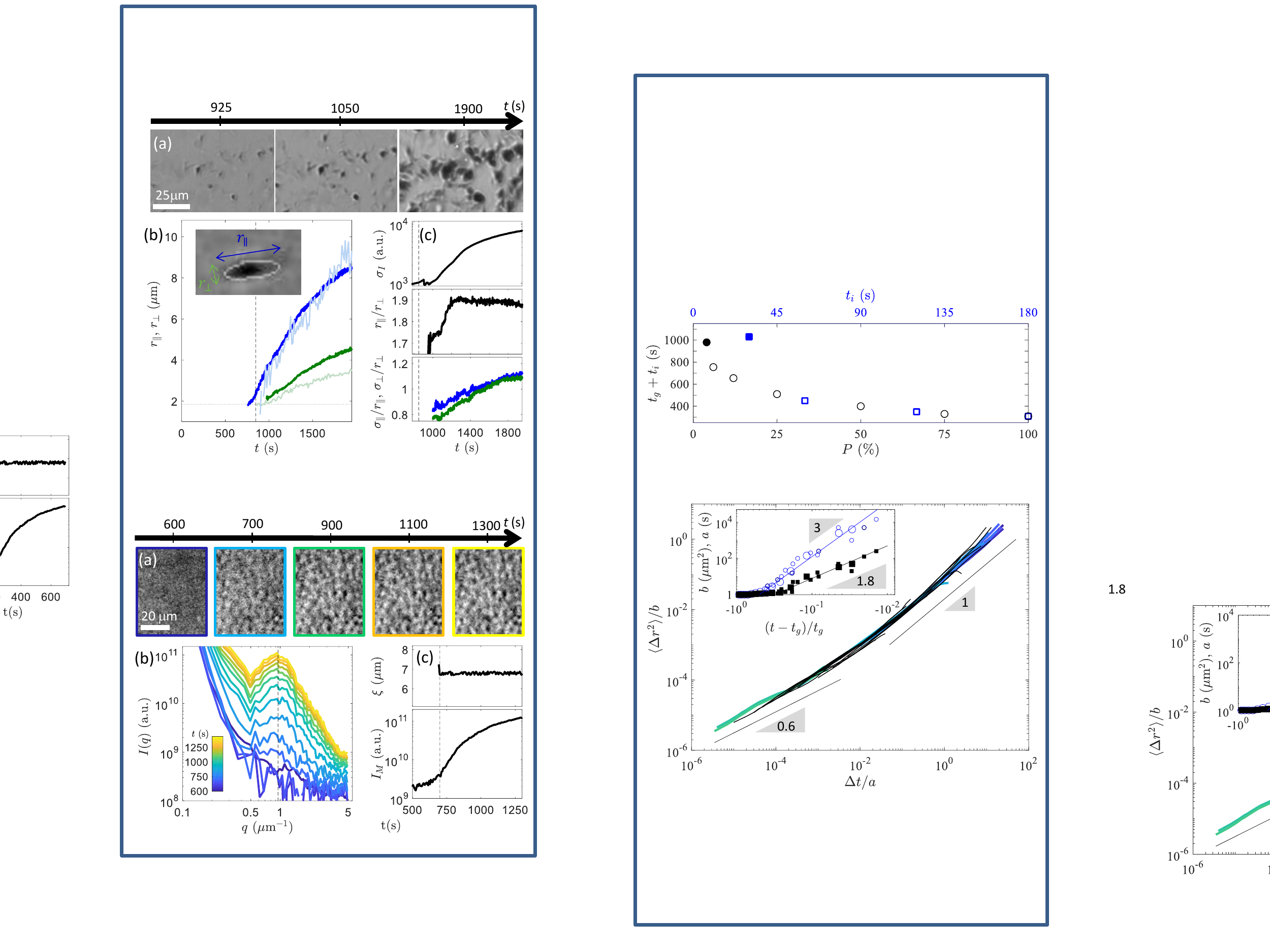}
      \caption{{\bf Gelation -- scaling behavior of the mean square displacement $\langle \Delta r^2 \rangle$ for $t<t_g$}. (a) $\langle \Delta r^2 \rangle$ as a function of $\Delta t$ rescaled by $b$ and $a$ respectively. The colored rescaled series correspond to the data presented in Fig.\ref{fig:microrheo}b with the same color code. The black lines correspond to rescaled data for other photo-activation conditions ($P,t_i$), namely: (6\%, 180~s), (12\%, 180~s) and (25\%, 180~s). Inset -- Evolution of the scaling parameters $a$ (circle) and $b$ (square) as a function of the normalized time $(t-t_g)/t_g$. $t_g$ is defined by $t_g=t^*$. Large symbols correspond to the photo-activation conditions $t_i=120$~s and $P=100\%$ and the smaller symbols correspond to the other photo-activation conditions. Lines are power law fits to the data: $a\sim [(t-t_g)/t_g]^y$ with $y=3.0 \pm 0.3$ and $b\sim [(t-t_g)/t_g]^z$ with $z=1.8 \pm 0.2$.
      }
    \label{fig:pregel}
\end{figure}

The time-resolved microrheology data displayed in Fig.~\ref{fig:microrheo}b are reminiscent of Larsen and Furst work on gelation~\cite{larsen2008}. These authors have indeed studied  the viscoelastic properties of physical and chemical polymer gels and focused on the sol-gel transition using time resolve microrheology. They showed that the mean square displacements of their colloidal tracers embedded in the dispersion could be scaled on a master curve. This is also the case for our data provided that $t<t_g$. As shown in Fig.\ref{fig:pregel}, the master curves is constructed by shifting individual $\langle \Delta r^2 \rangle$ curves along the ordinate and abscissa. Starting from the initial diffusive dynamics, subsequent $\langle \Delta r^2 \rangle$ curves are divided by a time-shift factor $a$ and a mean square displacement shift factor $b$. The resulting master curve has a logarithmic slope $\alpha=d\ln(\langle \Delta r^2 \rangle)/d\ln(\Delta t)=$1 for $\Delta t/a>1$, which decreases monotonically to a value $\alpha\simeq 0.6$ with decreasing $\Delta t/a$. At this point, subsequent curves can no longer be scaled onto this initial master curve.

For $t>t_g$, our data cannot be mapped on a master curve due to the dynamical heterogeneities displayed by the colloidal tracers.

Such dynamics, involving hopping processes, is reminiscent of behaviours monitored  with more complex soft solids like food gels mainly composed of proteins, polysaccharides and lipids~\cite{moschakis2013} or laponite gel/glasses~\cite{oppong2008}. Those dynamical heterogeneities have been inputted to micro structural features in the gel on length scales comparable to the probe particle sizes~\cite{gardel2005,rich2011}. Here, this is exactly the case: the colloid of diameter 5~$\mu$m and the gel correlation length $\xi\simeq 7$~$\mu$m are about the same size. These sponge-like network structure creates a gel backbone of relatively high supra-molecular density and regions where the supra-molecules have been depleted, resulting in a porous network.

Based on our observations and those of previous authors, we propose the following scenario for the evolution of the microstructure and the micro-rheology data for the gelation conditions, $P=100$~\% during $t_i=120$~s. In the early stage of the kinetics, the dispersion is homogeneous at the microscopic scale: the macro molecules are randomly distributed throughout the medium and diffusing freely. However, as the macro molecules diffuse, they move closer  and  get bound to one another through non covalent attractive interactions. These macro molecules chains grow extensively over time and eventually some span the container: this percolation characterizes the bulk gelation time. The hallmark of the gel point in rheology is defined by Chambon and Winter as the time when both moduli $G'$ and $G''$ exhibit approximately the same power-law frequency dependence $G' \sim G'' \sim f^{n}$ where $n=0.5$~\cite{chambon1987}. Moreover, near the gel point, the longest relaxation time and creep compliance are expected to exhibit power-law behavior so that $a\sim [(t-t_g)/t_g]^y$ and $b\sim [(t-t_g)/t_g]^z$ with $n=y/z$~\cite{adolf1990}. For our sample, the gel point is at $t=t_g \simeq 700$~s with $n=0.6$, a slightly higher value than the one proposed proposed by the model in~\cite{chambon1987}. The measured critical exponents $y$ and $z$ are respectively equal to $y=3.0 \pm 0.3$ and $z=1.8 \pm 0.2$ and lead to $n=0.60\pm 0.12$. Those values are self-consistent and informs us about the gelation mechanism. Indeed, our critical exponents are compatible with de Gennes Stauffer percolation model for gelation~\cite{de1976,stauffer1976} which leads to $y=2.69$ and $z=1.94$ but incompatible with Rousse or Zimm theory~\cite{martin1988} which leads respectively to ($y=4,z=2.67$) and ($y=2.67,z=2.67$).

For $t>700~s$, the structure is set and forms a sponge-like network with a correlation length, $\xi\simeq7$~$\mu$m. This is not the case for the  mechanical properties. Indeed the sample is still mostly viscous and the colloidal probes diffuse with only slight hindrance. As the microstructure densify and becomes more elastic for $t=700-1100$~s, the sample is a viscoelastic fluid: the colloidal probe show dynamical heterogeneities and the colloid is trapped within the structure at intermediate time scale but diffuse on long time scales. At the end of the kinetics, for $t>1200$~s the gel backbone is sufficiently elastic to hinder dynamical heterogeneity and the colloid probe the gel elasticity namely $G'\simeq 0.3$~Pa at relatively low frequencies, $f\simeq 0.1$~Hz.

\section{Discussion and conclusion}

\begin{figure}
	\centering
    \includegraphics[width=8cm]{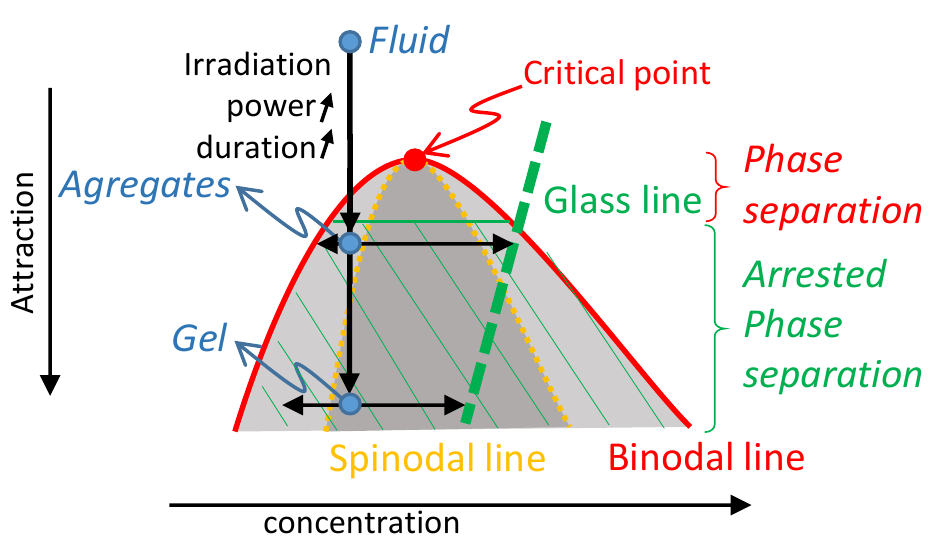}
      \caption{{\bf Schematic state diagram illustrating the physical pathway leading to aggregation and gelation}. State diagram as a function of the attraction between the building blocks and the concentration of the building blocks. The red line corresponds to the binodale line, the orange line to the spinodal line, the dash green line to the glass line, the horizontal green line to the separation between the phase separation and the arrested phase separation. The black arrows correspond to the physical pathway taken by the system and leading from a fluid state to the aggregates or the gel state.
      }
    \label{fig:sketch}
\end{figure}

In summary, we have shown that depending on photo-activation conditions, starting from a well dispersed suspension in fluid state, we can tune the self-assembly of viologen based molecular building blocks into aggregates or gels. Those results are compatible with a physical pathway that involves an interplay between phase separation and dynamical arrest. We hypothesize that the irradiation conditions allows to tune the attraction intensity between the building blocks as shown in Fig.~\ref{fig:sketch}. In other words, long and high power irradiation of the sample leads to the formation of interchain dimers causing a reinforcement of the interactions.

Gelation is a commonly observed phenomena in polymers and in supramoculer assemblies ~\cite{brinker2013}. It occurs for example when the molecular building blocks aggregate to form a
space spanning network. In our system gelation occurs for large irradiation power or irradiation, namely when the attraction between the building block becomes strong. The gel formed is very peculiar as it displays a very large correlation length $\xi$, a few order of magnitude larger than the size of the individual building blocks. Such a very large correlation length $\xi$ vouch for a very specific gelation mechanism: an arrested phase separation. In this scenario, the very large correlation length $\xi$ arises from density fluctuations in the vicinity of a critical point. At sufficiently high attraction and concentration in $\pi$-dimers, the dispersion undergo spinodal decomposition leading to bicountinuous network composed of a dilute and a concentrated fluid phase. Usually the spinodal decomposition proceeds and leads to the coexistence of two homogeneous fluid phases  separated by an interface. There is however a sweet spot where the glass transition mingle with the spinodal decomposition and leads to gelation, Fig.~\ref{fig:sketch}. In this scenario, the concentrated phase becomes denser and meets the glass line, it becomes kinetically arrested and freezes the phase separation process leading to a gel with a large correlation length $\xi$ \cite{gibaud2009, gibaud2012}.  

In our system, aggregation occurs for low irradiation power or time, namely when the attraction between the building blocks is mild. At the molecular level it might be explained by the lower concentration of $\pi$-dimers leading to supramolecular structures \textbf{3} or \textbf{4} having less inter-molecular interactions (Fig. \ref{fig:mechanism}). As shown in Fig.~\ref{fig:sketch}, this quench lead us in a region of the state diagram embedded between the binodal line and the spinodal line. In this region, phase separation proceed through the formation of isolated droplets composed of a dense fluid in background compose of a  dilute fluid phase. Usually the phase separation proceeds and leads to the coexistence of two homogeneous phase fluid separated by an interface. Here, as for the arrested spinodal decompositon, as the concentration of the droplet increases and reach the glass line, the phase separation is stopped leading to isolate aggregate as observed experimentally. 

We also observe that the aggregates display a spindle-like structures. This structure is typical of liquid-crystal dispersion and corresponds to the hallmark of the Isotropic-Nematic transition~\cite{modlinska2015, gibaud2017}. The Isotropic-Nematic transition is a phase separation and therefore is compatible with the arrested phase separation scenario. It however deals with liquid-crystal molecular systems; i.e. rod-shaped molecules. The spindle-like structures are called tactoids and correspond to Nematic droplets composed of elongated molecules that tend to point in the same direction but do not have any positional order. The spindle-like shape results from the molecules orientation and the interplay between the interfacial tension and the splay and bend elastic constants of the nematic domain~\cite{tortora2011}. In viologen-based molecules, liquid-crystal phase have recently been observed~\cite{casella2014, casella2016, pibiri2019}. However tactoids have not been reported maybe due to their transient nature. We tried to check for a birefringent signature of our spindle-like structure under cross-polarisation microscopy but fail due to the opacity of the structure. At this point, our best guess is that the molecules self-assemble into liquid-crystal due to the photo-induced formation of \textbf{3} presenting an appropriate rod-shaped structure. Following this hypothesis, these more rigid structures could then form tactoids due to inter-molecular interactions with adjacent boxes by orbital overlap involving more than two viologen radical-cations as already reported \cite{wang2016, qi2016}. Metal-metal interaction between palladium complexes might also be at stake.

To conclude, thanks to time resolved experiments such as the microstructure analysis via microscopy and the mechanical properties analysis via microrheology, we were able to propose a unique physical pathway leading to both aggregation and gelation. This pathway is driven by an arrested phase separation. Such a scenario is new to supramolecular systems but has already been observed in colloidal dispersions \cite{lu2008}, polymers \cite{glassman2015} and proteins solutions \cite{cardinaux2007}. It opens up new perspectives in the field of viologen-based responsive supramolecular gels \cite{datta2017, dhiman2020, wei2017, xu2018, Zhang2013}. This work provides a physical analysis of a highly complex chemical system. The chemical evolution in terms of interactions and reactions pathway remains to be elucidated and compared to the physical pathway proposed in this article. Indeed the influence of many parameters remains to be addressed and will be the subject of a subsequent article: exact structures of supramolecular self-assemblies (box, folded chain, inter-chain $\pi$-dimers), nature of the supramolecular interactions involved (orbital overlap of viologen radical-cations, metal-metal bonds), solubility and electrostatic effects of the charged building blocks.

\section*{Acknowledgements}
The authors thank the Ecole Normale Supérieure de Lyon (ENSL) and the Centre National de la Recherche Scientifique (CNRS) for financial, logistical and administrative supports. C. Roizard thanks ILM and IDEXLYON for the COLUMN internship grant. This work was supported by the LABEX iMUST (ANR-10-LABX-0064) of Université de Lyon, within the program "Investissements d'Avenir" (ANR-11-IDEX-0007) operated by the French National
Research Agency (ANR).


%

\end{document}